\documentclass[aps,pre,twocolumn,showpacs]{revtex4}
\usepackage{amssymb,amsmath}
\usepackage[dvips]{graphicx}

\begin{document}

\title{Chimera Ising Walls in Forced Nonlocally Coupled Oscillators}

\author{Yoji Kawamura}\email{kawamura@ton.scphys.kyoto-u.ac.jp}
\affiliation{Department of Physics, Graduate School of Sciences,
Kyoto University, Kyoto 606-8502, Japan}

%\date{\today}
\date{December 12, 2006}

\pacs{05.45.Xt, 82.40.Ck}

%%%%% abstract
\begin{abstract}
  Nonlocally coupled oscillator systems can exhibit an exotic
  spatiotemporal structure called {\it chimera}, where the system
  splits into two groups of oscillators with sharp boundaries,
  one of which is phase-locked and the other is phase-randomized.
  Two examples of the chimera states are known: the first one appears
  in a ring of phase oscillators, and the second one is associated
  with the two-dimensional rotating spiral waves.
  In this article, we report yet another example of the chimera state
  that is associated with the so-called Ising walls in one-dimensional
  spatially extended systems, which is exhibited by a nonlocally
  coupled complex Ginzburg-Landau equation with external forcing.
  %%
  %%We demonstrate the existence of chimera Ising walls in nonlocally
  %%coupled complex Ginzburg-Landau equation with external forcing.
\end{abstract}

\maketitle

%%%%% section 1
\section{Introduction} \label{sec:intro}

Nonlocally coupled oscillator systems can exhibit a remarkable class of
patterns called {\it chimera}, in which identical oscillators separate sharply
into two domains, one coherent and phase locked, the other incoherent and
drifting~\cite{ref:kuramoto02,ref:shima04,ref:abrams04,ref:kuramoto06}.
These two groups of the oscillators together maintain stable organized patterns
in the system.
The existence of such patterns was first noticed and explained in a ring of phase
oscillators~\cite{ref:kuramoto02}, and studied in further detail~\cite{ref:abrams04}.
Furthermore, it was found that the chimera states also appear in rotating spiral
waves in two-dimensional spatially extended systems~\cite{ref:shima04}.
Recently, an interesting pattern similar to the chimera states studied in
Refs.~\cite{ref:kuramoto02,ref:abrams04} was also found in nonlocally
coupled Hodgkin-Huxley equations with excitatory and inhibitory synaptic coupling,
where such patterns spontaneously appears as a result of the instability of the
uniform state, in contrast to the original chimera that appears when the uniform
state is stable~\cite{ref:sakaguchi06}.

In this paper, we present another simple example of the chimera state
associated with the Ising walls in one-dimensional spatially extended systems,
which is exhibited by a {\it nonlocally} coupled complex Ginzburg-Landau
equation (CGLE) {\it with} a parametric forcing.
Recently, a {\it nonlocally} coupled CGLE {\it without} a forcing has been
studied intensively~\cite{ref:kuramoto95,ref:kuramoto97,ref:kuramoto98,
  ref:nakao99,ref:tanaka03,ref:manrubia03,ref:casagrande05},
and a {\it locally} coupled CGLE {\it with} a forcing has also been
investigated widely~\cite{ref:coullet90,ref:mizuguchi93,ref:ohta97,
  ref:battogtokh00,ref:aranson02,ref:teramoto04,ref:kobayashi06}.
To our knowledge, Battogtokh considered the {\it nonlocally} coupled CGLE
{\it with} the forcing for the first time, and demonstrated various interesting
phenomena, e.g., nonequilibrium Ising-Bloch transitions~\cite{ref:battogtokh}.
Here, we will focus only on the chimera Ising wall, which is the simplest one
among them, but its theoretical analysis has not yet been carried out.

The organization of the present paper is the following.
In Sec.~\ref{sec:chimera}, we introduce our model and illustrate its normal
and chimera Ising walls by numerical simulations.
In Sec.~\ref{sec:phase}, we reduce our model to the phase model, and numerically
demonstrate that the reduced phase model also exhibits the chimera Ising walls.
In Sec.~\ref{sec:theory}, a functional self-consistency equation is derived by
introducing a space-dependent order parameter, and its numerical solution is
compared with the numerical simulation presented in Sec.~\ref{sec:phase}.
Concluding remarks will be given in the final section.

%%%%% section 2
\section{Chimera Ising Walls} \label{sec:chimera}

We consider the following equation that describes a system of nonlocally
coupled limit-cycle oscillators driven by a parametric external forcing,
%%% eq.1
\begin{align}
\partial_t A&=\left(1+i c_0\right)A
-\left(1+i c_2\right)\left|A\right|^2 A \nonumber \\
&+K\left(1+i c_1\right)\left(B-A\right)+\gamma A^{\ast},
\label{eq:cgl}
\end{align}
which we call a forced nonlocally coupled complex Ginzburg-Landau equation.
Here $c_0, c_1, c_2, K$, and $\gamma$ are real parameters, a complex amplitude
$A(x,t)$ represents the state of a local limit-cycle oscillator at location $x$
and time $t$, and $A^{\ast}$ is the complex conjugation of $A$.
The quantity $B(x,t)$ represents the nonlocal coupling defined by
%%% eq.2
\begin{equation}
B\left(x,t\right)=\int dx'\,
G\left(x-x'\right)A\left(x',t\right),
\label{eq:mean}
\end{equation}
%%% eq.3
\begin{equation}
%G\left(x\right)=\frac{1}{2}e^{-\left|x\right|},
G\left(x\right)=\frac{1}{2}\exp\left(-\left|x\right|\right),
\label{eq:nonlocal}
\end{equation}
where the nonlocal coupling function $G(x)$ is normalized in the infinite domain.
$K$ represents the coupling strength, and $c_1$ is the phase shift of
the coupling~\cite{ref:kuramoto95,ref:kuramoto97,ref:kuramoto98,
  ref:nakao99,ref:tanaka03,ref:manrubia03,ref:casagrande05}.
The last term $\gamma A^{\ast}$ represents the effect of the parametric external forcing
with almost double the natural frequency, whose intensity is given by $\gamma$,
and $c_0-c_2$ stands for the frequency mismatch~\cite{ref:coullet90,ref:mizuguchi93,
  ref:ohta97,ref:battogtokh00,ref:aranson02,ref:teramoto04,ref:kobayashi06}.
In the absence of the coupling and the external forcing, i.e., $K=\gamma=0$,
Eq.~(\ref{eq:cgl}) is simply given by $\partial_t A=(1+i c_0)A-(1+i c_2)|A|^2 A$,
which is the simplest limit-cycle oscillator called Stuart-Landau oscillator
~\cite{ref:kuramoto84,ref:pikovsky01}, so that Eq.~(\ref{eq:cgl}) describes
a system of forced nonlocally coupled oscillators.
In addition, Eq.~(\ref{eq:cgl}) is a normal form that can be derived from
a wide class of reaction-diffusion systems near the Hopf bifurcation point
under particular assumptions by using the center-manifold reduction method
~\cite{ref:kuramoto84,ref:kuramoto95,ref:tanaka03,ref:coullet90}.

%In our numerical simulations, the continuous media was replaced with a long
%array of $N=2048$ oscillators with sufficiently small and fixed separation
%$\varDelta x=0.005$ between the neighboring oscillators, and the Neumann
%boundary condition (zero-flux) was imposed.
In our numerical simulations, the continuous media of size $L$ was discretized using
$N=2048$ grid points with sufficiently small and fixed separation $\varDelta x=0.005$,
i.e., $L=N\varDelta x$, and the Neumann boundary condition (zero-flux) was imposed.
Our numerical results are unchanged if we further increase the number of grid points $N$
or the system size $L$.
The initial condition was such that $A(x<0,t=0)=1.0$ and $A(x\geq 0,t=0)=-1.0$.
In what follows, we fix the parameter values as $c_0=1.0$, $c_1=-0.5$, and $c_2=1.0$.
Note that we set the frequency mismatch at zero, i.e., $c_0-c_2=0$,
for the sake of simplicity.
Furthermore, we fix the ratio of the coupling strength to the force
intensity as $K/\gamma=1.5$.

\begin{figure}
\centering
\includegraphics[width=6cm]{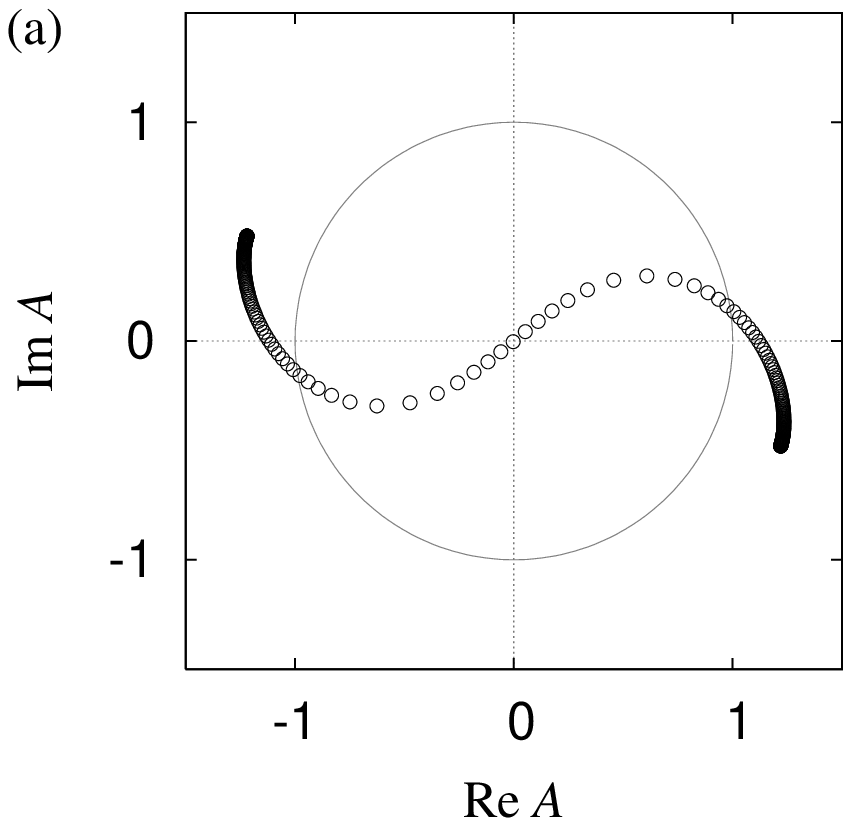}
\includegraphics[width=8cm]{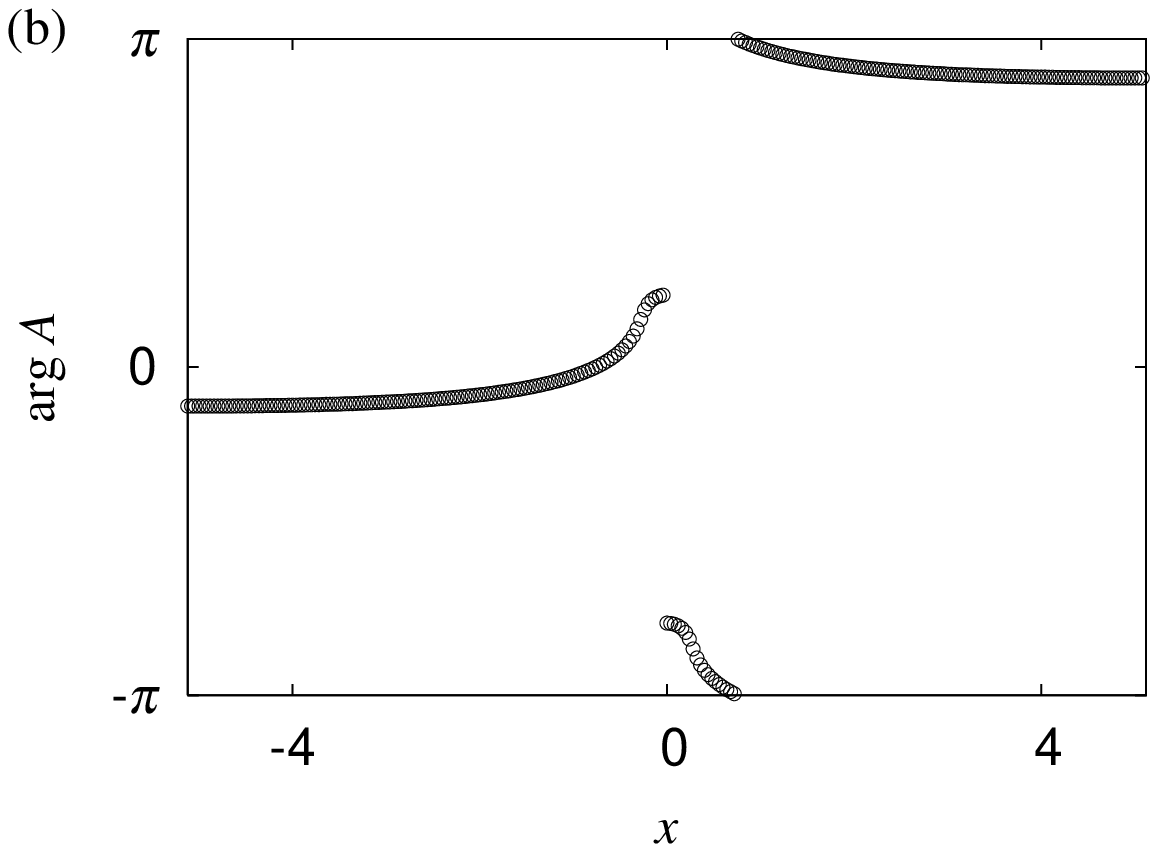}
\includegraphics[width=8cm]{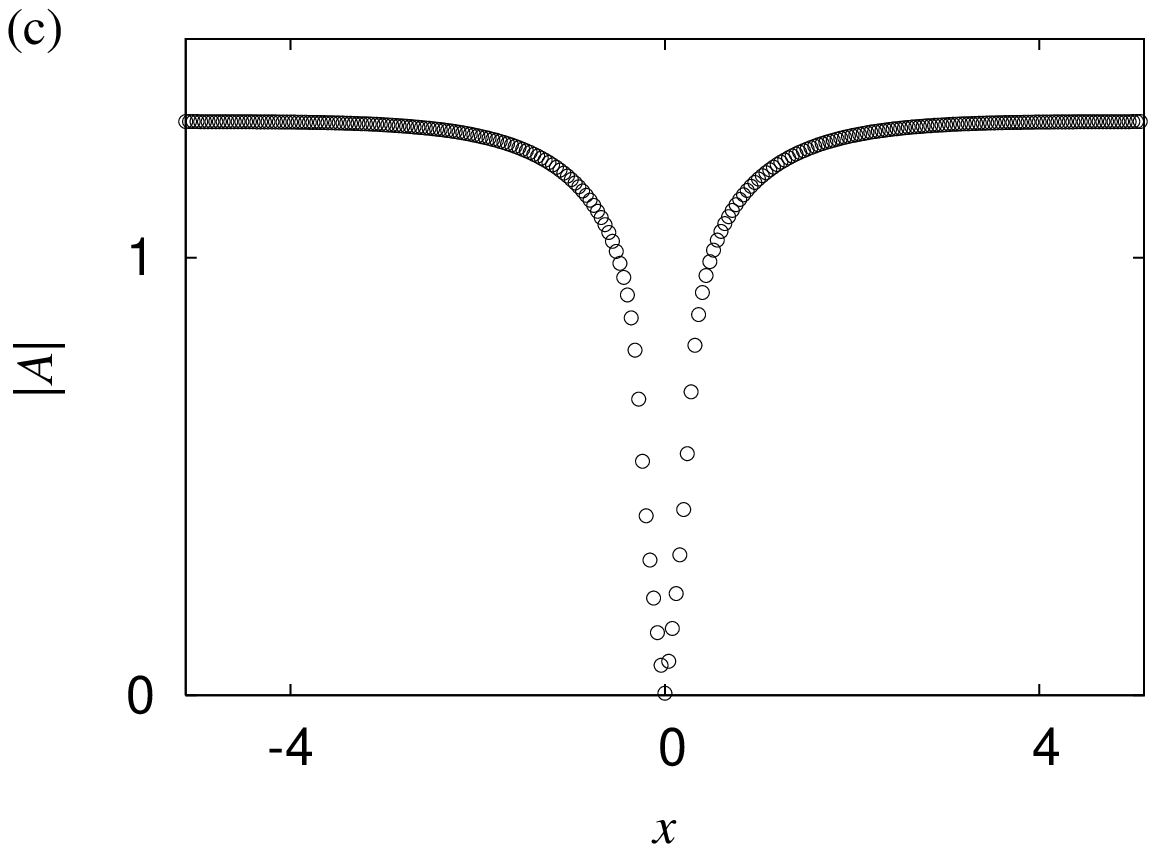}
\caption{Phase portrait (a), spatial phase profile (b),
  and spatial modulus profile (c) of the normal Ising wall
  exhibited by the forced nonlocally coupled complex
  Ginzburg-Landau equation (\ref{eq:cgl}) in the strong
  coupling case, $K=1.5$ and $\gamma=1.0$.
  The numerical data are represented by the open circles
  (only 1 in every 8 oscillators is plotted).
  The limit-cycle orbit $|A|=1$ of the local oscillator is also
  displayed in (a).}
\label{fig:normal}
\end{figure}

Figure \ref{fig:normal} displays a phase portrait (a), a spatial profile of the
phase (b), and a spatial profile of the modulus (c) for the strong coupling case,
$K=1.5$ and $\gamma=1.0$, obtained by a direct numerical simulation of
Eq.~(\ref{eq:cgl}).
The phase portrait is given by a set of grid points in the complex plane
each representing the state of a local oscillator at a given time.
It is found that a normal Ising wall appears and all local oscillators are
completely phase-locked.

\begin{figure}
\centering
\includegraphics[width=6cm]{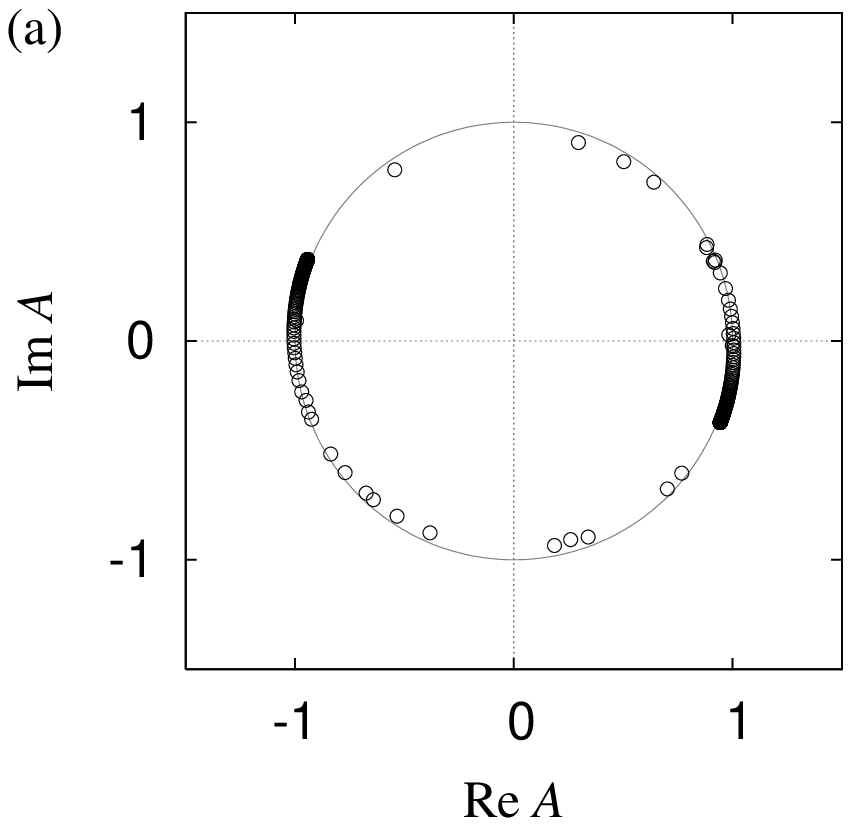}
\includegraphics[width=8cm]{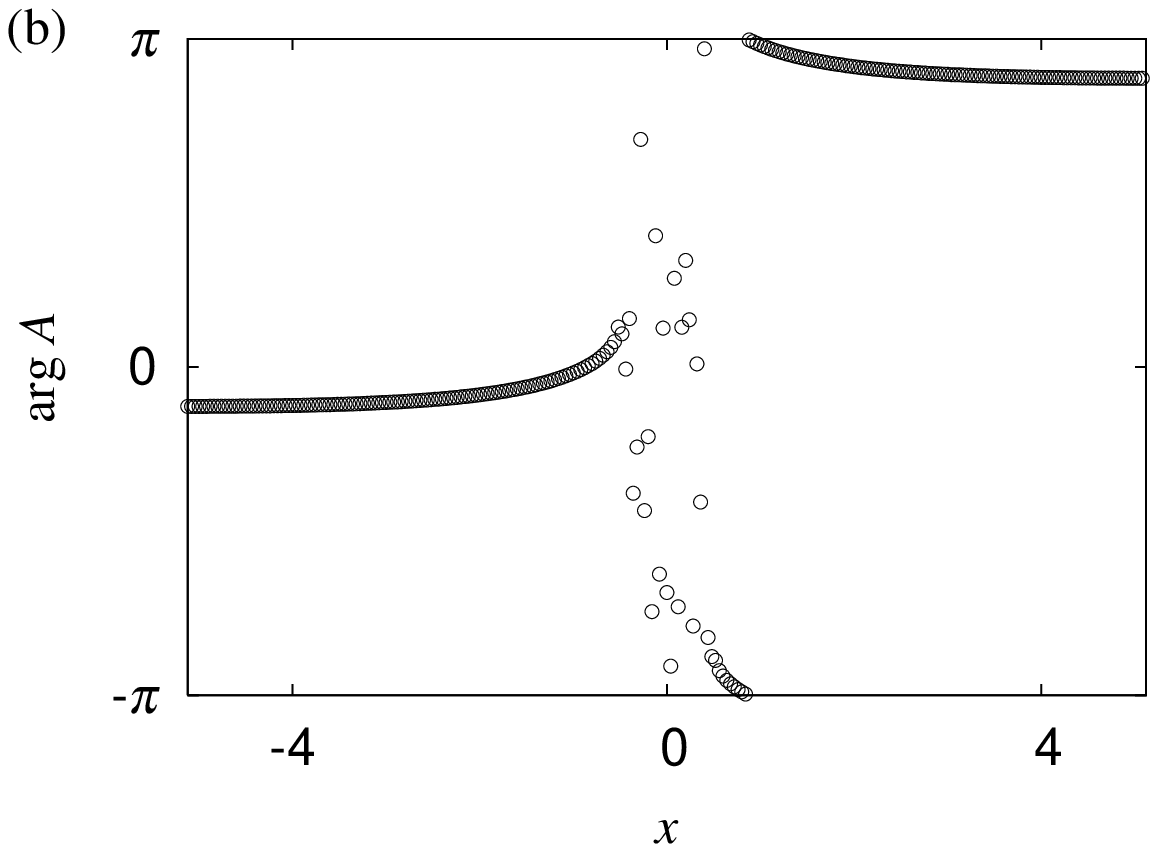}
\includegraphics[width=8cm]{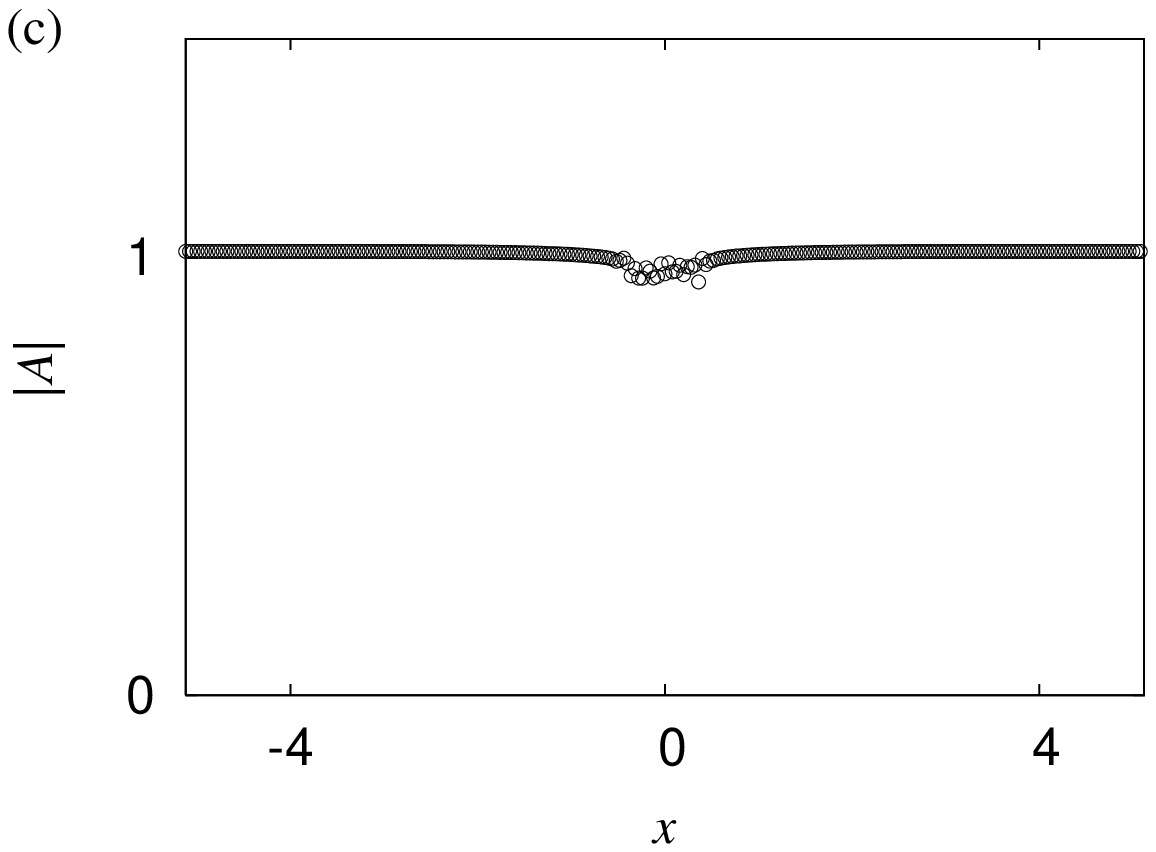}
\caption{Phase portrait (a), spatial phase profile (b),
  and spatial modulus profile (c) of the chimera Ising wall
  exhibited by the forced nonlocally coupled complex
  Ginzburg-Landau equation (\ref{eq:cgl}) in the the weak
  coupling case, $K=0.06$ and $\gamma=0.04$.
  The numerical data are represented by the open circles
  (only 1 in every 8 oscillators is plotted).
  The limit-cycle orbit $|A|=1$ of the local oscillator is also
  shown in (a).}
\label{fig:chimera}
\end{figure}

Figure \ref{fig:chimera} displays a phase portrait (a), a spatial profile of
the phase (b), and a spatial profile of the modulus (c) for the weak coupling case,
$K=0.06$ and $\gamma=0.04$.
It is found that a chimera Ising wall appears, which consists of incoherent drifting
oscillators near the center ($x=0$) and coherent phase-locked oscillators in the
peripheral regions.
All the oscillators take almost the full amplitude, $|A|=1$, so that the system
has no phase singularity.
At the same time, the spatial continuity of the pattern near the center ($x=0$)
is lost, so that a pair of local oscillators infinitely close to each other in
this region are not always close in the state space.
%%% fig.1
%%% fig.2

We can estimate the critical coupling strength below which the oscillator at
the center of the normal Ising wall starts to drift incoherently, leading to
the chimera state.
From the spatial symmetry of the normal Ising wall, the values of $A$ and $B$
vanish at the center, i.e., $A(x=0,t)=B(x=0,t)=0$.
If we regard $B$ as an external forcing \cite{ref:kuramoto06},
%%which corresponds to the mean-field picture,
the linearized equation for the complex amplitude $A(x=0,t)$ at the
center is given by
%%% eq.4
\begin{equation}
\partial_t A=\left(1+i c_0\right)A-K\left(1+i c_1\right)A+\gamma A^{\ast}.
\label{eq:wall}
\end{equation}
A linear stability analysis of the stationary solution $A(x=0,t)=0$
gives the following eigenvalues:
%%% eq.5
\begin{equation}
\lambda_\pm=1-K\pm\sqrt{\gamma^2-\left(c_0-Kc_1\right)^2}.
\label{eq:wall}
\end{equation}
Therefore, the necessary condition for the appearance of the chimera Ising wall,
i.e., the condition for the oscillator at the center ($x=0$) to drift, is
expressed as $\lambda_{+}>0$.  In fact, when $K<1$ is satisfied under our
parameter conditions, the oscillator at the center ($x=0$) starts to drift and a
chimera Ising wall appears.
Hereafter, we focus on the chimera Ising walls and consider the situation where
the coupling strength is much smaller than 1, i.e., $K\ll 1$.

%%%%% section 3
\section{Phase reduction} \label{sec:phase}

In order to investigate the nature and the origin of the chimera Ising walls in
further detail, we reduce Eq.~(\ref{eq:cgl}) to a phase equation for the phase
$\phi(x,t)$, which is much easier to analyze.  When the coupling strength $K$
and the forcing intensity $\gamma$ are sufficiently small, the phase reduction
method \cite{ref:winfree80,ref:kuramoto84,ref:pikovsky01} is applicable,
which is actually the case for our weak coupling case, $K=0.06$ and
$\gamma=0.04$. The reduced equation takes the form
%%% eq.6
\begin{align}
\partial_t\phi=&-K\sigma\int dx'\,G\left(x-x'\right)
\Bigl[\sin\bigl(\phi-\phi'+\alpha\bigr)-\sin\alpha\Bigr] \nonumber \\
&-\gamma\tau\sin\bigl(2\phi+\beta\bigr),
\label{eq:reduction}
\end{align}
where the condition $c_0 - c_2 = 0$ was used.
Here $\phi'$ is the abbreviation of $\phi(x',t)$.
The new parameters are related to the original parameters through
%%% eq.7,8
\begin{align}
%\sigma e^{i\alpha}&=(1-i c_1)(1+i c_2), \label{eq:parameter1} \\
%\tau e^{i\beta}&=(1+i c_2), \label{eq:parameter2}
\sigma \exp\left(i\alpha\right)&=(1-i c_1)(1+i c_2), \label{eq:parameter1} \\
\tau \exp\left(i\beta\right)&=(1+i c_2), \label{eq:parameter2}
\end{align}
where $\sigma$, $\tau$, $\alpha$, and $\beta$ are real.
When the time scale is changed so that the force intensity
is normalized, $t\to t/\gamma\tau$, Eq.~(\ref{eq:reduction})
can be rewritten as
%%% eq.9
\begin{align}
\partial_t\phi=&-K_\gamma\int dx'\,G\left(x-x'\right)
\Bigl[\sin\bigl(\phi-\phi'+\alpha\bigr)-\sin\alpha\Bigr] \nonumber \\
&-\sin\bigl(2\phi+\beta\bigr),
\label{eq:phase}
\end{align}
where
%%% eq.10
\begin{equation}
K_\gamma=\frac{K\sigma}{\gamma\tau}.
\label{eq:coupling}
\end{equation}

\begin{figure}
\centering
\includegraphics[width=8cm]{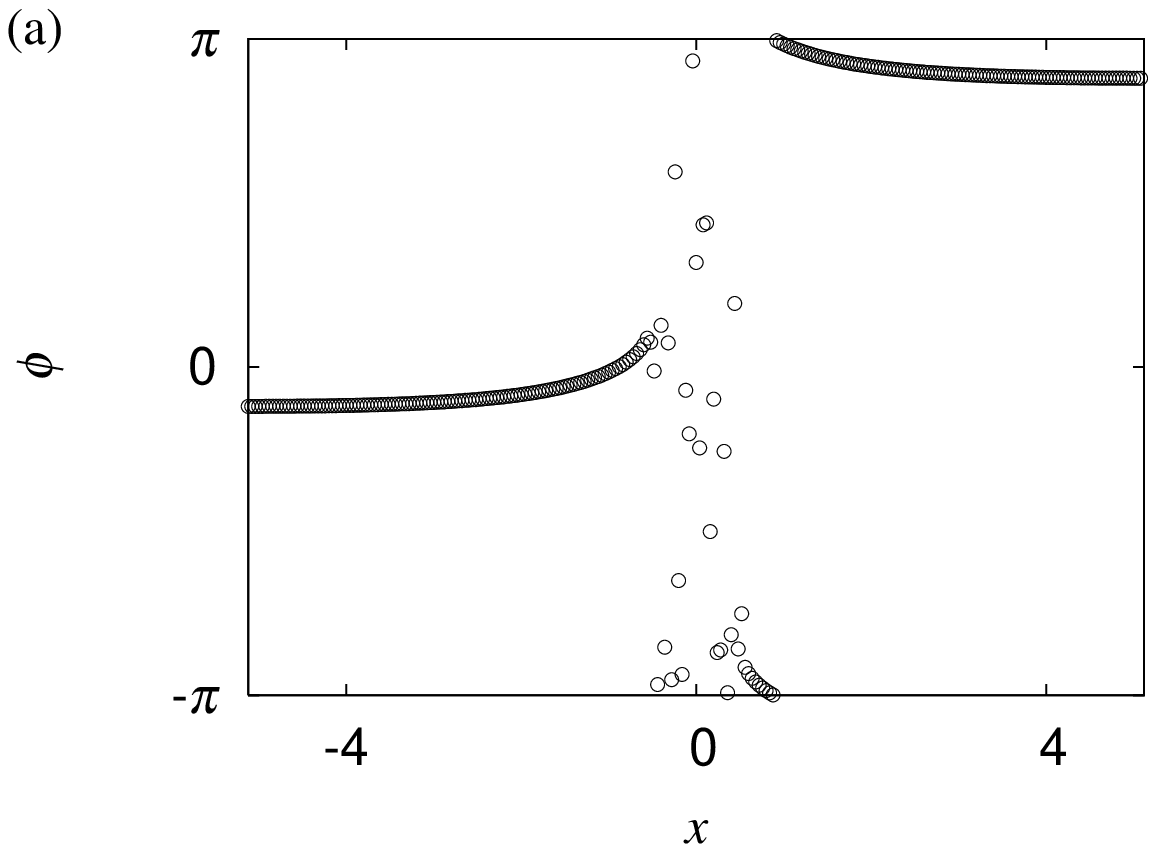}
\includegraphics[width=8cm]{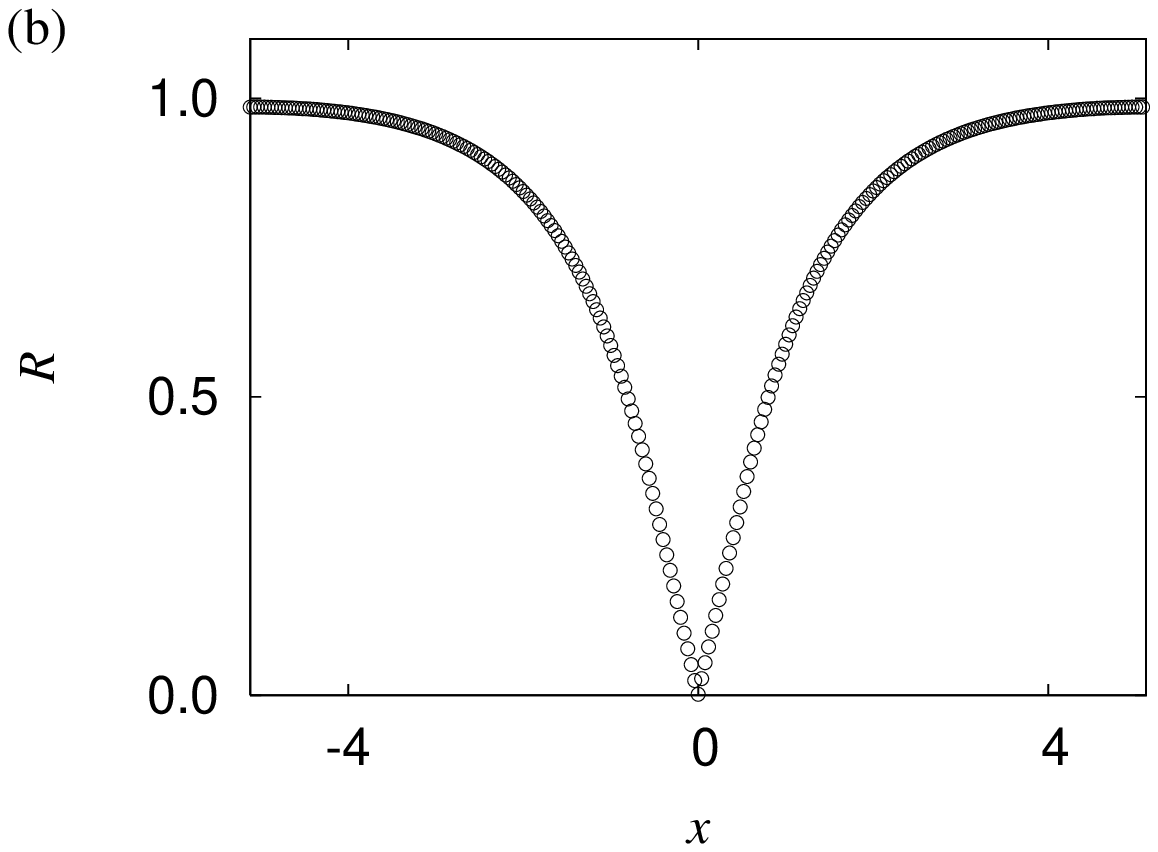}
\includegraphics[width=8cm]{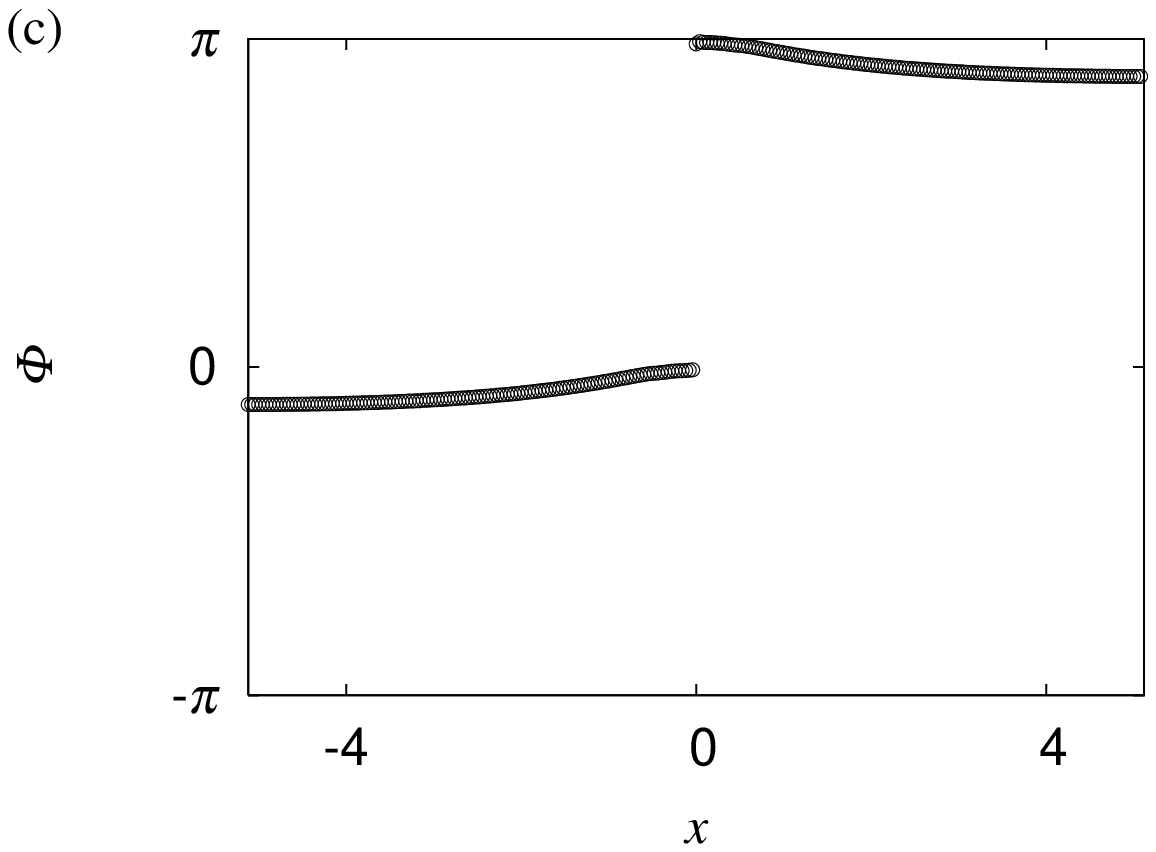}
\includegraphics[width=8cm]{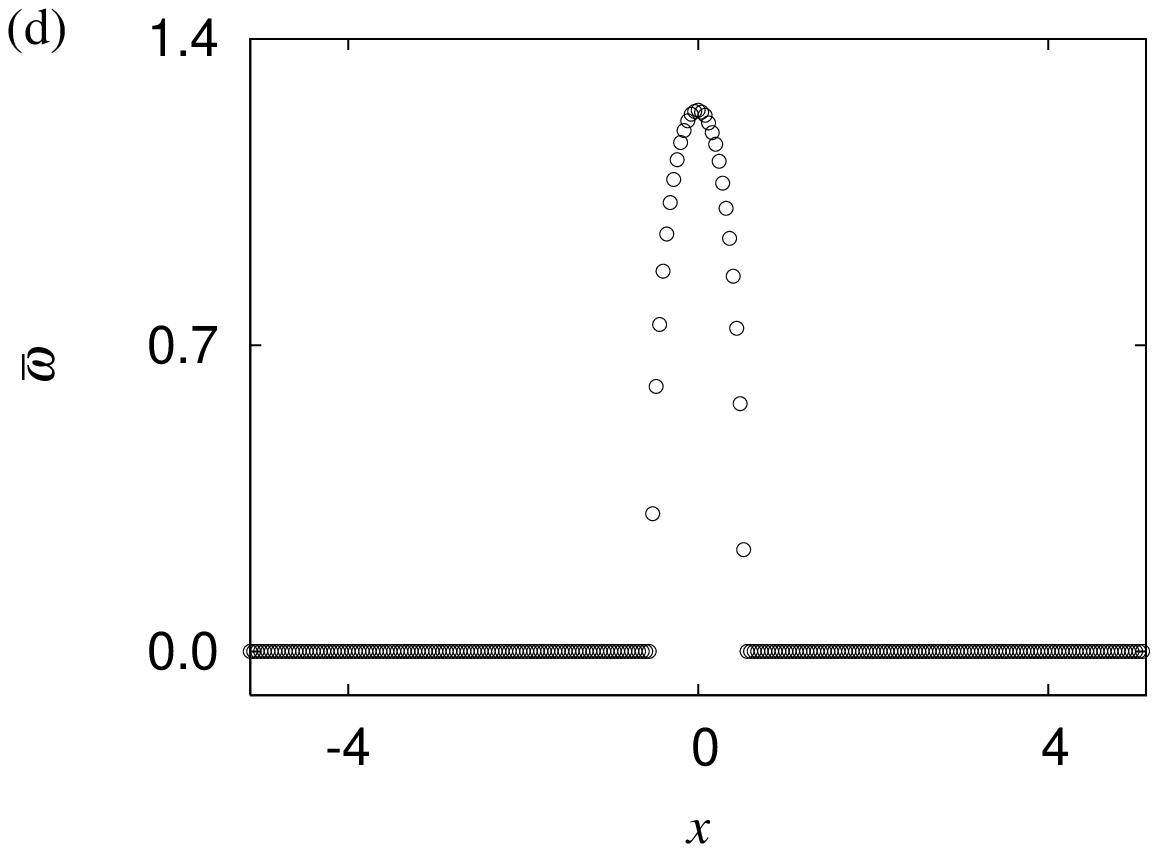}
\caption{Spatial profiles of the local oscillator phase $\phi$ (a),
  the order-parameter modulus $R$ (b), the order-parameter phase
  $\Phi$ (c), and the mean frequency $\bar{\omega}$ of the local
  oscillators (d), obtained by a numerical simulation of the phase
  equation (\ref{eq:phase}). The numerical data are represented by
  open circles (only 1 in every 8 oscillators is plotted).
  \label{fig:phase}}
\end{figure}

Figure~\ref{fig:phase}(a) displays a spatial profile of the phase $\phi$ of the
local oscillators obtained by a numerical simulation of Eq.~(\ref{eq:phase})
using parameter conditions corresponding to Fig.~\ref{fig:chimera}. The phase pattern
is very similar to that in Fig.~\ref{fig:chimera}(b). As in the previous case,
only the local oscillators near the center ($x=0$) are drifting incoherently,
and all other oscillators are phase-locked.

Now let us introduce a space-dependent complex order parameter with
modulus $R(x)$ and phase $\Phi(x)$ through
%%% eq.11
\begin{equation}
%R\left(x\right)e^{i\Phi\left(x\right)}=
%\int dx'\,G\left(x-x'\right)e^{i\phi\left(x',t\right)},
R\left(x\right)\exp\bigl(i\Phi\left(x\right)\bigr)=
\int dx'\,G\left(x-x'\right)\exp\bigl(i\phi\left(x',t\right)\bigr).
\label{eq:order}
\end{equation}
We assume the order parameter to be time independent, which will be
confirmed below. In terms of this order parameter, Eq.~(\ref{eq:phase})
may be expressed in the form of a single-oscillator equation
%%% eq.12
\begin{equation}
\partial_t\phi=
-K_\gamma\Bigl[R\sin\bigl(\phi-\Phi+\alpha\bigr)-\sin\alpha\Bigr]
-\sin\bigl(2\phi+\beta\bigr),
\label{eq:one}
\end{equation}
or, if we further introduce a space-dependent
{\it effective force function} $H_x(\phi)$ through
%%% eq.13
\begin{equation}
H_x\left(\phi\right)=K_\gamma
\Bigl[R\sin\bigl(\phi-\Phi+\alpha\bigr)-\sin\alpha\Bigr]
+\sin\bigl(2\phi+\beta\bigr),
\label{eq:function}
\end{equation}
Eq.~(\ref{eq:phase}) can be expressed as
%%% eq.14
\begin{equation}
\partial_t\phi=-H_x\left(\phi\right).
\label{eq:basic}
\end{equation}
Our effective force function is composed of the {\it internal} force
(the first harmonic term) and the {\it external} force (the second harmonic term),
which is similar to the order function introduced by Daido
~\cite{ref:daido97,ref:strogatz00,ref:acebron05}.
%%
%% In this case, we need the order parameter only for the first
%% harmonic of the phase coupling function because the second harmonic
%% term represents not a coupling but an external forcing. However one
%% may call $H_x(\phi)$ a space-dependent order function.

The spatial profiles of $R(x)$ and $\Phi(x)$ obtained by the
numerical simulation of Eq.~(\ref{eq:phase}) are displayed in
Fig.~\ref{fig:phase}(b) and Fig.~\ref{fig:phase}(c), respectively.
We can see that the order-parameter modulus has a vanishing value at the center,
and both the modulus and the phase of the order parameter are time independent.
The distribution of the mean frequency $\bar{\omega}(x)$ of the local oscillators,
which is defined by a long-time average of $\partial_t \phi$, is also displayed in
Fig.~\ref{fig:phase}(d). In the phase-locked domain, the oscillation frequencies
are identically zero, while in the drifting domain they are distributed.
%%% fig.3

\begin{figure}
\centering
\includegraphics[width=8cm]{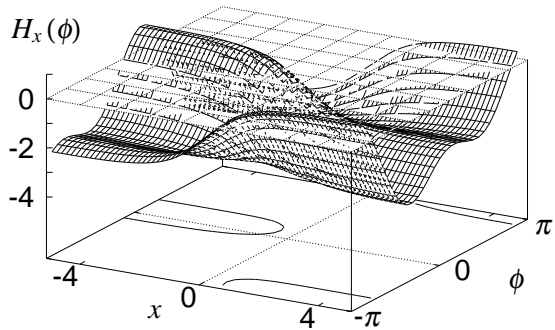}
\caption{Dependence of the effective force function $H_x(\phi)$
  on $x$ and $\phi$.
  The contour lines corresponding to $H_x(\phi)=0$ are also
  displayed in the base plane.
  Note the gap near the center $(x=0)$, where $H_x(\phi)$ does not
  cross the zero plane.}
\label{fig:function}
\end{figure}

The dependence of the effective force function $H_x(\phi)$ on $x$ and
$\phi$ is displayed in Fig.~\ref{fig:function}. It is found that $H_x(\phi)$
crosses the zero plane only twice in every $2\pi$ interval of $\phi$ at each $x$
in the peripheral regions, which will make our analysis simple. We should also
note that $H_x(\phi)$ is always below zero near the center ($x=0$), namely,
$H_x(\phi)$ never crosses the zero plane in this region.
%%% fig.4
In the next section, we develop a self-consistent theory for determining
the spatial profiles of the quantities displayed in Fig.~\ref{fig:phase}.

%%%%% section 4
\section{Self-Consistent Theory} \label{sec:theory}

We now develop a self-consistent theory that can reproduce
our simulation results by generalizing the earlier theories
~\cite{ref:kuramoto02,ref:shima04,ref:abrams04,ref:kuramoto06,ref:winfree80,
  ref:kuramoto84,ref:pikovsky01,ref:daido97,ref:strogatz00,ref:acebron05}.
There are two possible cases regarding the solution of Eq.~(\ref{eq:basic}).
Let $H_{\rm min}(x)$ and $H_{\rm max}(x)$ denote the minimum and the maximum of
$H_x(\phi)$ in every $2\pi$ interval of $\phi$ at each $x$, respectively.
Then, the two cases are expressed by
%%% eq.15
\begin{equation}
H_{\rm min}\left(x\right)<0<H_{\rm max}\left(x\right)
\quad\text{(case I)},
\label{eq:case1}
\end{equation}
and
%%% eq.16
\begin{equation}
H_{\rm min}\left(x\right)>0\quad \mbox{or} \quad
0>H_{\rm max}\left(x\right)\quad\text{(case II)}.
\label{eq:case2}
\end{equation}
Correspondingly, the oscillators are divided into two groups.
In the case I, which corresponds to the group of the phase-locked
oscillators in the peripheral regions, Eq.~(\ref{eq:basic}) admits
only one pair of stable and unstable fixed points.
We denote the stable fixed point formally as
%%% eq.17
\begin{equation}
\phi_0\left(x\right)=H_x^{-1}\left(0\right),
\label{eq:phi0}
\end{equation}
where $H_x^{-1}$ is the inverse function of $H_x$.
The average frequency $\bar{\omega}(x)$ of the oscillators in this group
is identically zero,
%%% eq.18
\begin{equation}
\bar{\omega}\left(x\right)=0.
\label{eq:omega1}
\end{equation}

The case II corresponds to the group of the drifting oscillators,
for which Eq.~(\ref{eq:basic}) admits a drifting solution.
The average frequency $\bar{\omega}(x)$ is formally expressed as
%%% eq.19
\begin{equation}
\bar{\omega}\left(x\right)=2\pi\left[\int_0^{2\pi}
\frac{d\phi}{-H_x\left(\phi\right)}\right]^{-1},
\label{eq:omega2}
\end{equation}
which depends on $x$.
The contribution to the order parameter from the drifting oscillators can
be computed by the standard method~\cite{ref:kuramoto02,ref:kuramoto84}.
That is, we use the invariant measure, i.e., the probability
density $p_x(\phi)$ associated with the drift motion.
Noting that the probability density for the phase at $x$ to take
on value $\phi$ must be inversely proportional to the drift velocity
given by the right hand side of Eq.~(\ref{eq:basic}), we obtain
%%% eq.20
\begin{equation}
p_x\left(\phi\right)=
C_x\left[\;-H_x\left(\phi\right)\right]^{-1},
\label{eq:px}
\end{equation}
where $C_x$ is the normalization constant given by
%%% eq.21
\begin{equation}
C_x=\left[\int_0^{2\pi}\frac{d\phi}
{-H_x\left(\phi\right)}\right]^{-1}.
\label{eq:cx}
\end{equation}

\begin{figure}[h]
\centering
\includegraphics[width=8cm]{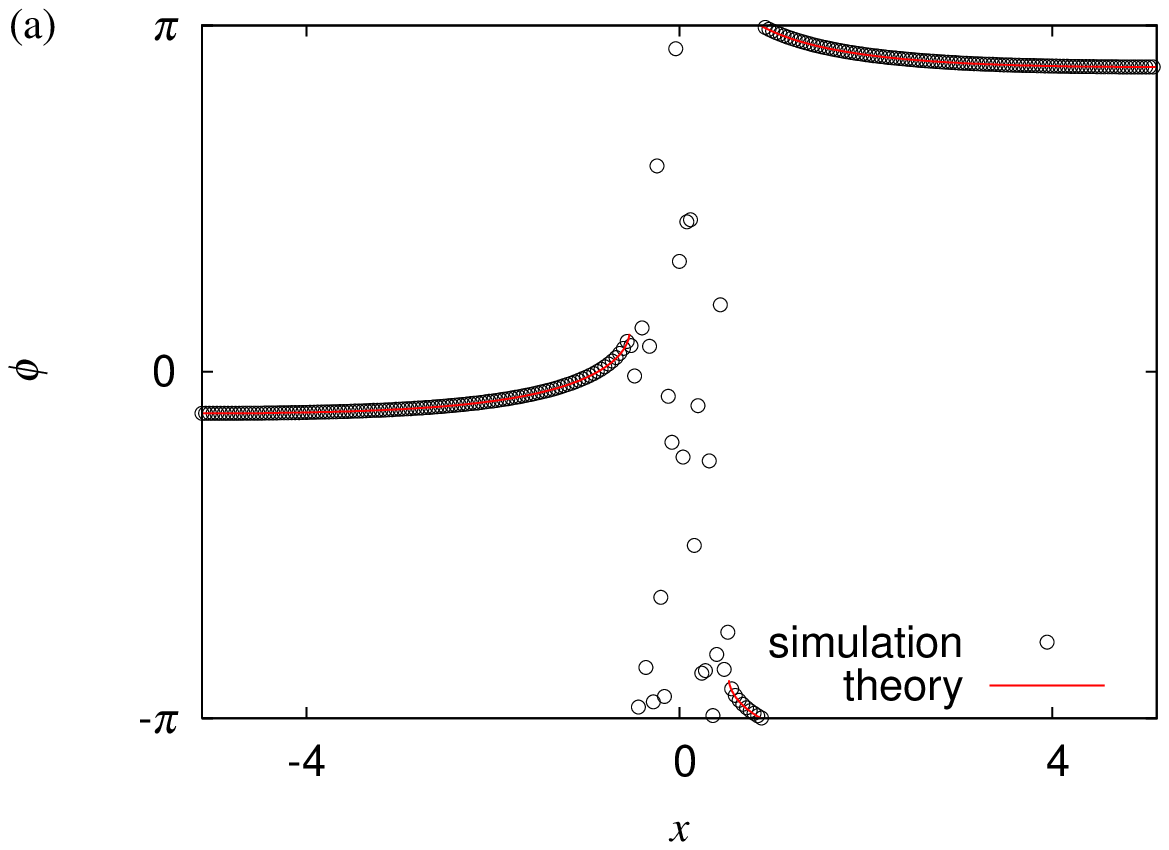}
\includegraphics[width=8cm]{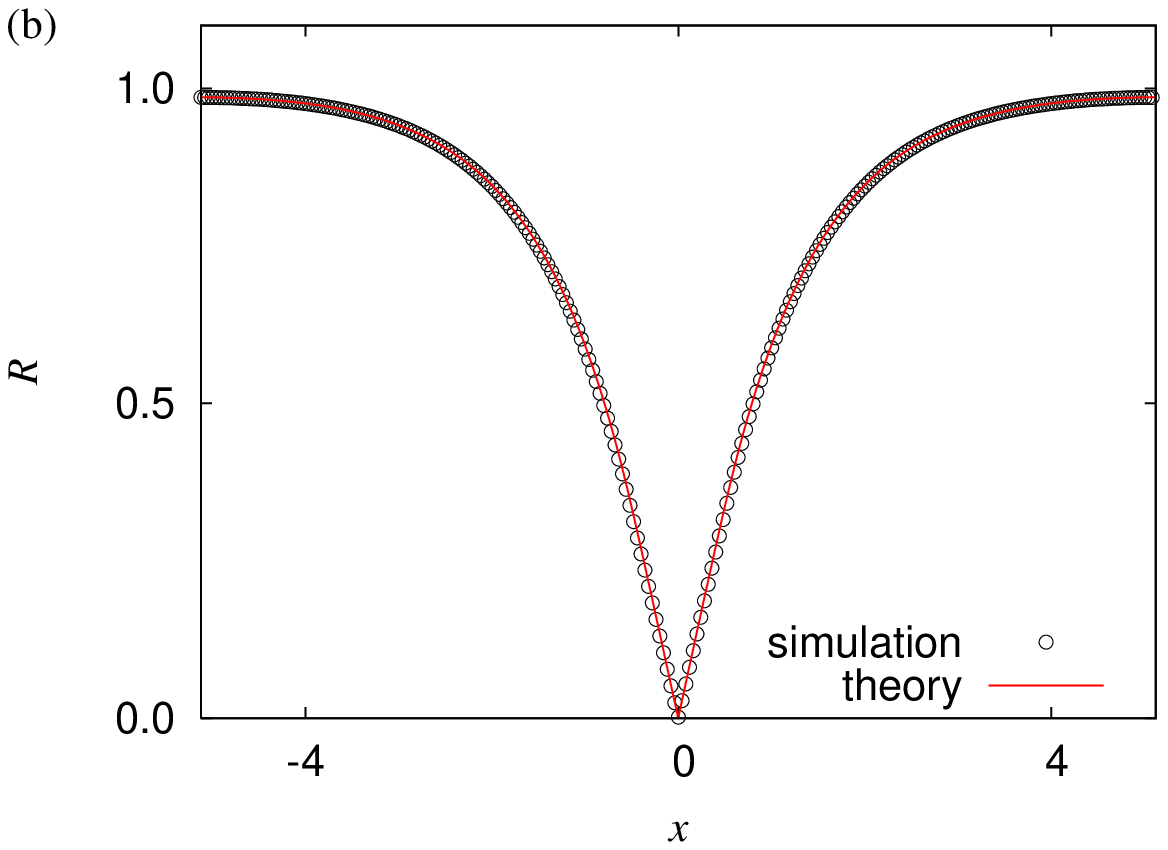}
\includegraphics[width=8cm]{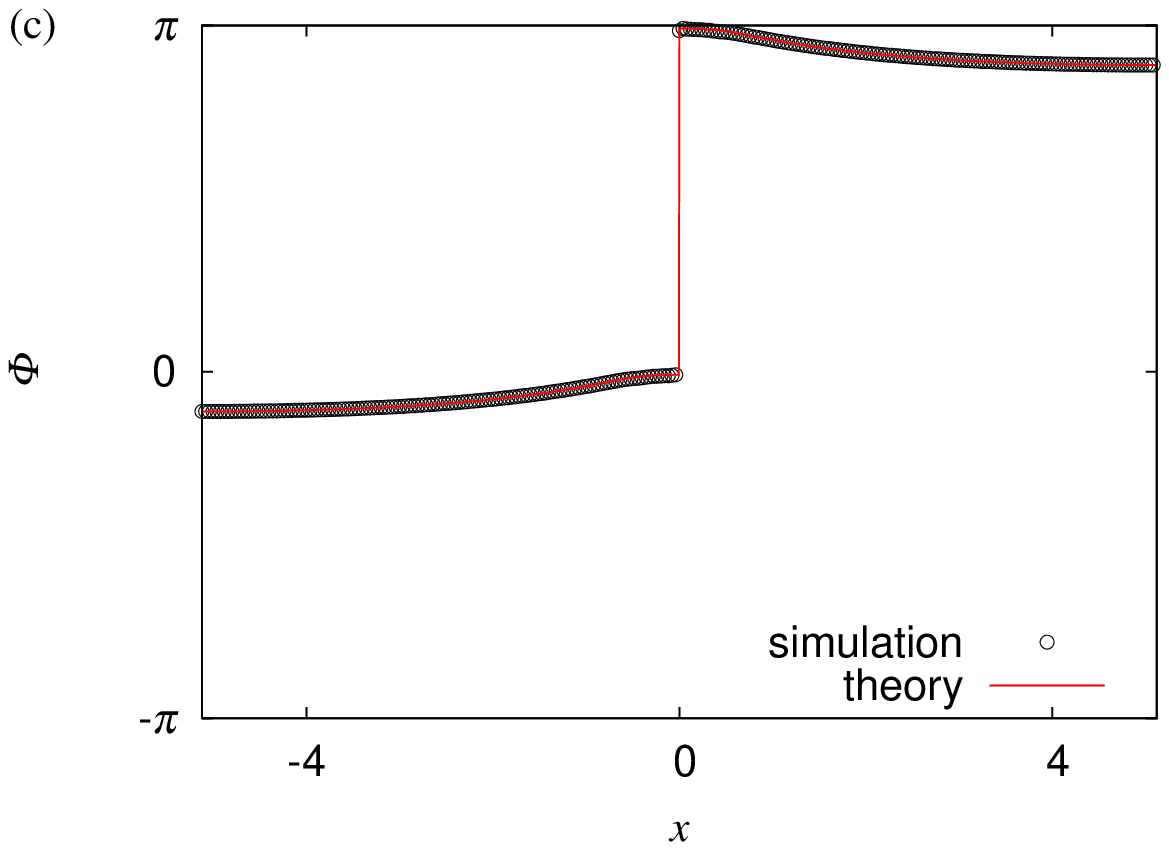}
\includegraphics[width=8cm]{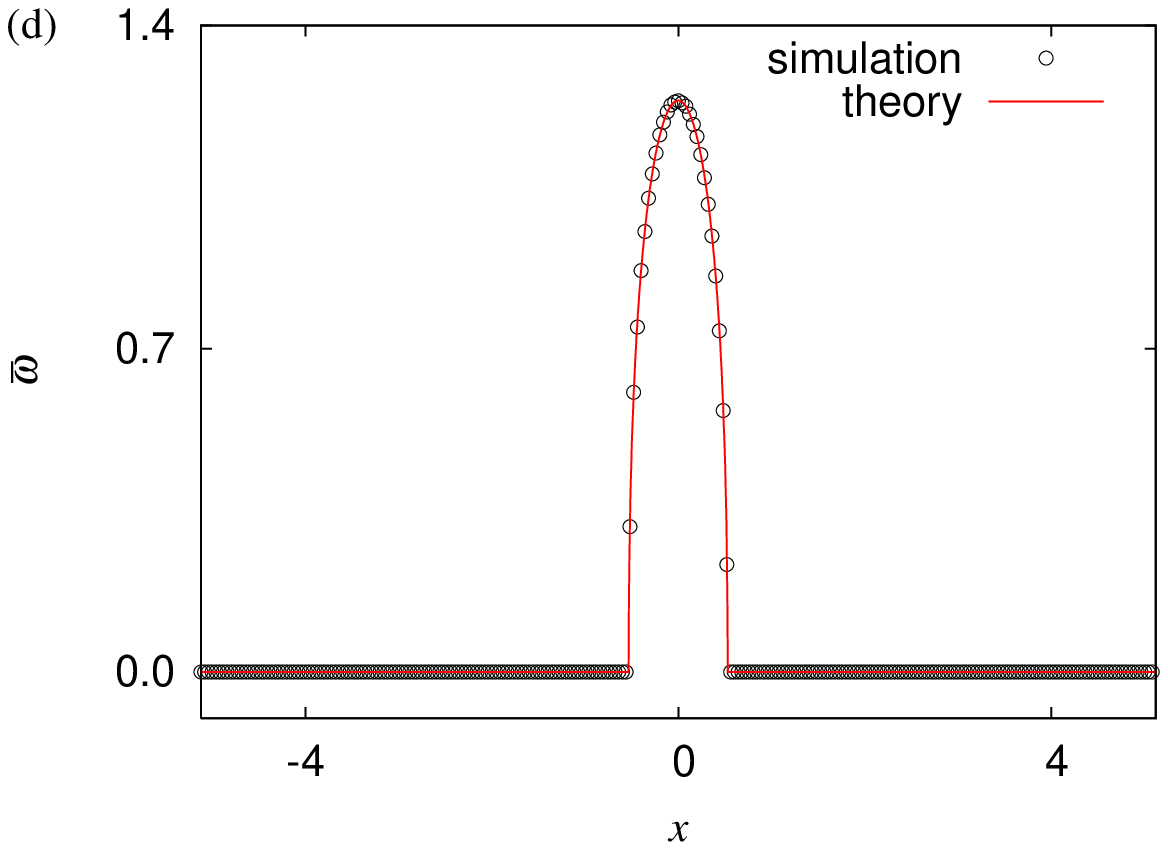}
\caption{(Color online) Comparison between the theory and the
  simulation results.
  The open circles and the solid lines are the simulation data
  and the theoretical curves, respectively.
  Spatial profiles of the local oscillator phase $\phi$ (a),
  the order-parameter modulus $R$ (b), the order-parameter phase $\Phi$ (c),
  and the mean frequency $\bar{\omega}$ of the local oscillators (d).}
\label{fig:theory}
\end{figure}

Putting together the two types of contributions to the order parameter,
we finally obtain a functional self-consistency equation in the form
%%% eq.22
\begin{equation}
%R\left(x\right)e^{i\Phi\left(x\right)}=
R\left(x\right)\exp\bigl(i\Phi\left(x\right)\bigr)=
\int dx'\,G\left(x-x'\right)h\left(x'\right),
\label{eq:self}
\end{equation}
where
%%% eq.23
\begin{equation}
h\left(x\right)=
\begin{cases}
\displaystyle{e^{i\phi_0\left(x\right)}} &
\quad \text{(case I)}, \\
\displaystyle{\int_{0}^{2\pi}p_x\left(\phi\right)e^{i\phi}d\phi} &
\quad \text{(case II)}.
\end{cases}
\label{eq:hx}
\end{equation}
We can determine the quantities shown in Fig.~\ref{fig:phase} from this
functional self-consistency equation. The solution can be obtained numerically
by an iteration procedure, which is compared with the results of the numerical
simulation in Fig.~\ref{fig:theory}.
The agreement between the theory and the numerical simulation is excellent
for all quantities.
%%% fig.5

We finally make a brief comment. The order-parameter phase does not drift in
our chimera Ising walls, which makes our analysis easy because we do not need to
calculate the collective frequency. In the two early examples of the chimera
states, we had to work with a nonlinear eigenvalue problem, or to determine the
various quantities together with the collective frequency (i.e., the eigenvalue)
~\cite{ref:kuramoto02,ref:shima04}. \\

%%%%% section 5
\section{Concluding remarks} \label{sec:remarks}

We present a new type of the chimera states associated with the Ising walls in
the one-dimensional forced nonlocally coupled complex Ginzburg-Landau equation.
We have focused on the weak coupling case, where the phase reduction
method is applicable, and reduced the original model to the phase model.
Generalizing the previous theories, we derived a functional self-consistency
equation to be satisfied by the order parameter.
Its solution successfully reproduced our simulation results carried out
on this phase model.

The chimera state studied in Refs.~\cite{ref:kuramoto02,ref:abrams04} seems
rather special because the boundary effects are crucial~\cite{ref:kuramoto06}.
In contrast, the chimera Ising walls stably exist regardless of the boundary
effects, and they survive even in spatially infinitely extended systems, like
the chimera spiral waves that persist in two-dimensional spatially extended
systems~\cite{ref:shima04,ref:kuramoto06}.
Our preliminary analysis suggests that a {\it chimera hole} solution does
not seem to exist in a one-dimensional nonlocally coupled (cubic) complex
Ginzburg-Landau equation without a parametric forcing.
Therefore, our chimera Ising wall may be the simplest example of the chimera
states which is still relevant to the real-world phenomena.

%%%%% acknowledgments
\begin{acknowledgments}
The author is grateful to Y.~Kuramoto, D.~Battogtokh, H.~Nakao,
and S.~Shima for valuable discussions.
\end{acknowledgments}

%%%%% references

\end{document}